\documentclass[preprint]{rsl}

\usepackage[utf8]{inputenc}
\nocite{*}

\usepackage{multirow}
\usepackage[T1]{fontenc}
\usepackage{newtxtext,newtxmath}
\usepackage{xcolor}
\usepackage{svg}
\usepackage{float}
\usepackage{colortbl}

\svgsetup{inkscapelatex=false}
\usepackage[colorlinks=true,linkcolor=blue]{hyperref}
\usepackage{tikz}
\usepackage{pgfplots}
\usepackage{pgfplotstable}
\pgfplotsset{compat=1.7}
\usepackage{subcaption}
\usepgfplotslibrary{groupplots}
\usepackage{booktabs}

\usepackage{url}
\usepackage{xcolor}
\usepackage{lineno}

\title{Anomaly detection in radio galaxy data with\\ trainable COSFIRE filters}

\author{Steven Ndung'u, Trienko Grobler,  Stefan J. Wijnholds, and George Azzopardi}

\begin{document}

\maketitle

%
%

\begin{abstract}
Detecting anomalies in radio astronomy is challenging due to the vast amounts of data and the rarity of labeled anomalous examples. Addressing this challenge requires efficient methods capable of identifying unusual radio galaxy morphologies without relying on extensive supervision. This work introduces an innovative approach to anomaly detection based on morphological characteristics of the radio sources using trainable COSFIRE (Combination of Shifted Filter Responses) filters as an efficient alternative to complex deep learning methods. The framework integrates COSFIRE descriptors with an unsupervised Local Outlier Factor (LOF) algorithm to identify unusual radio galaxy morphologies. Evaluations on a radio galaxy benchmark data set demonstrate strong performance, with the COSFIRE-based approach achieving a geometric mean (G-Mean) score of 79\%, surpassing the 77\% achieved by a computationally intensive deep learning autoencoder. By characterizing normal patterns and detecting deviations, this semi-supervised methodology overcomes the need for anomalous examples in the training set, a major limitation of traditional supervised methods. This approach shows promise for next-generation radio telescopes, where fast processing and the ability to discover unknown phenomena are crucial.

\end{abstract}

\section{Introduction}
\label{introduction}
The advent of next-generation radio telescopes has ushered in an era of unprecedented data volumes in radio astronomy, with facilities like the Square Kilometre Array (SKA) expected to generate exabytes of data annually \cite{farnes2018science}. While processing such vast data sets poses significant challenges, it also presents unique opportunities for serendipitous discoveries of novel astrophysical phenomena. The systematic exploration of these data sets necessitates robust anomaly detection methodologies capable of identifying both rare, theoretically anticipated phenomena and entirely novel, unforeseen ones. 

\begin{figure}[t]
\centering
\footnotesize
\begin{tabular}{c@{\hspace{0.6cm}}cc}
\toprule
Typical Sources & Real Anomalies & Synthetic Anomalies \\
\midrule
    \includegraphics[width=0.20\columnwidth]{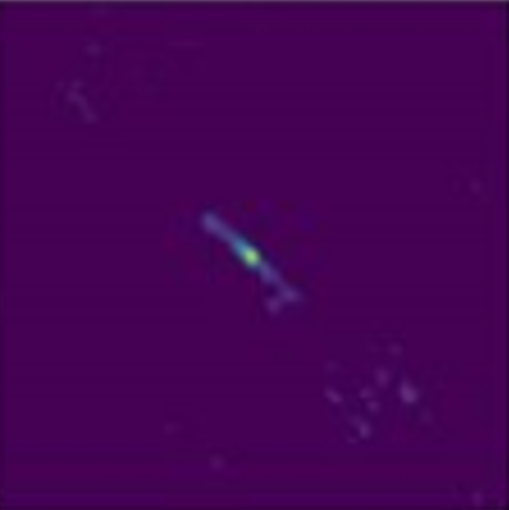} & 
    \includegraphics[width=0.20\columnwidth]{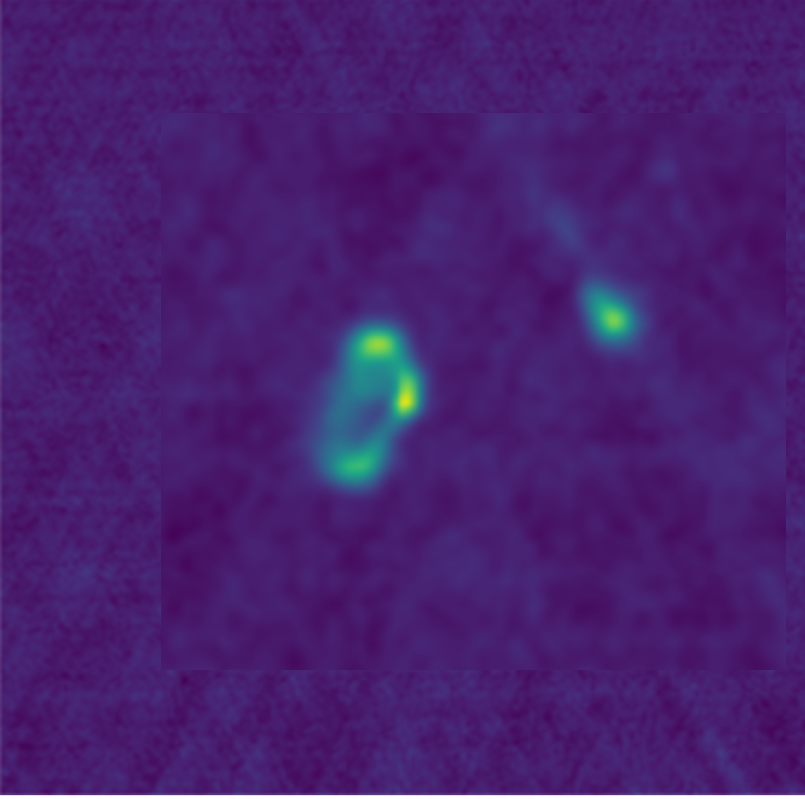} & 
    \includegraphics[width=0.20\columnwidth]{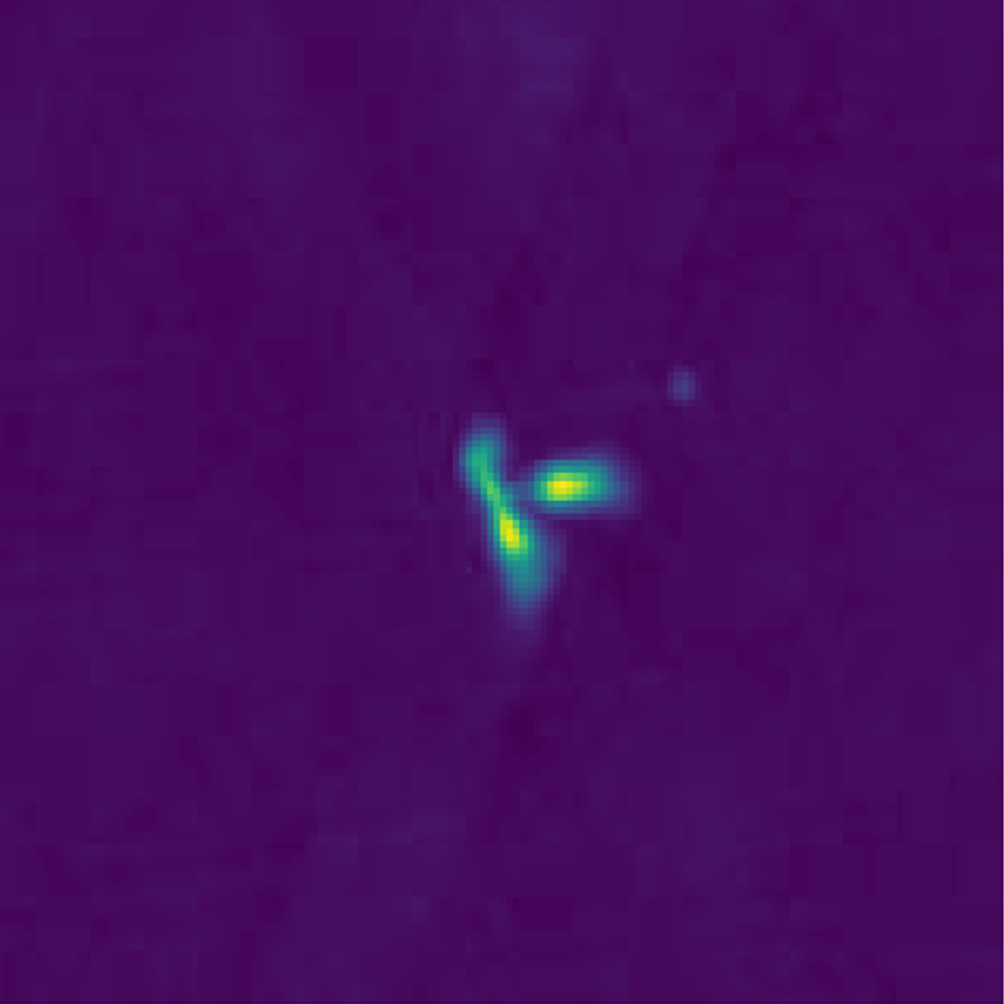} \\
FRI & RRG & Example 1 \\
\includegraphics[width=0.20\columnwidth]{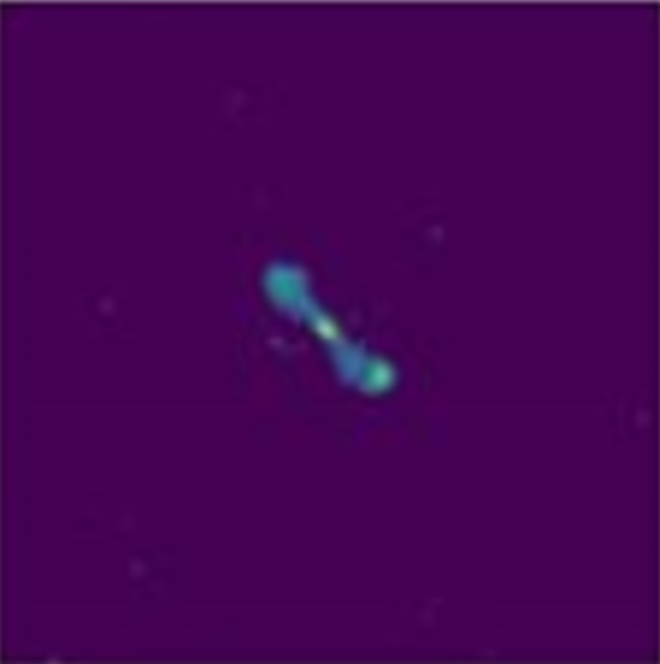} & 
\includegraphics[width=0.20\columnwidth]{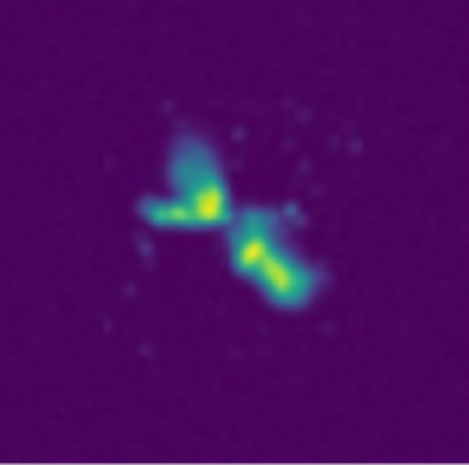} & 
\includegraphics[width=0.20\columnwidth]{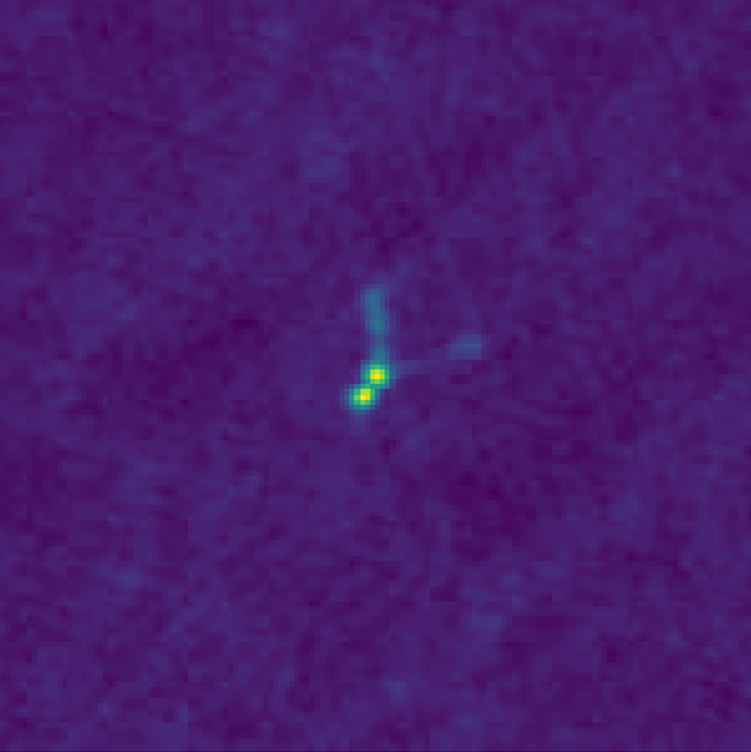} \\
FRII & XRG &   Example 2 \\
\bottomrule
\end{tabular}

\caption{Examples of typical radio galaxies and both real and synthetic anomalous radio sources: FRI (edge-darkened source), FRII (edge-brightened source), XRG (x-shaped source), RRG (ring-shaped source) and synthetic sources with non-standard morphologies.}
\label{fig: example_images}

\end{figure}

Machine learning approaches, particularly those focused on learning compact representations of typical radio sources, have emerged as promising tools for this task. Auto-encoders (AEs) have been successfully applied to anomaly detection across diverse astronomical data sets \cite{ventura_2022,brand2025cara}, but their high computational cost and opaque latent representations present significant drawbacks. To address these challenges, we use trainable COSFIRE filters \cite{azzopardi2012trainable}, which are efficient, explainable and rotation-tolerant, and whose responses form the basis of a feature descriptor \cite{10.1093/mnras/stae821}. COSFIRE filters are automatically configured to detect and describe the spatial arrangement of emission blobs—cores, jets, and lobes—that characterize a radio source’s morphology. This approach has shown notable efficacy, accuracy, and robustness in radio galaxy classification and image retrieval \cite{10.1093/mnras/stae821,10.1093/mnras/staf230} outperforming deep learning approaches in certain cases. Notably, COSFIRE's rotation invariance ensures that galaxies with the same intrinsic morphology receive consistent labels, regardless of their apparent orientation due to the telescope's line of sight. 
Figure~\ref{fig: example_images} shows examples of typical radio galaxies (FRI - Fanaroff–Riley Class I, FRII - Fanaroff–Riley Class II) \cite{fanaroff1974morphology}, real anomalous sources (XRG - X-shaped Radio Galaxy, and RRG - Ring Radio Galaxy) \cite{proctor2011morphological}, and synthetic anomalous  sources used in this work (see Section~\ref{datatab_section}).

\section{Data and Methodology}
\label{sec: data_and_methodology}
\subsection{Data set}
\label{datatab_section}

\begin{table}[h]
\footnotesize
\centering
\caption{Distribution of the FRGADB data set across training, validation, and test sets for each galaxy type. Note that XRG and RRG are used only in the test set.}
\label{original_dataset_distrb}
\begin{tabular}{@{}lcccc@{}}
\toprule
\textbf{Galaxy Type} & \textbf{Total Samples} & \textbf{Training} & \textbf{Validation} & \textbf{Test} \\
\midrule
\multicolumn{5}{@{}l}{\textit{Regular classes}} \\
FRI  & 211 & 137 & 22 & 22 \\
FRII & 523 & 387 & 53 & 53 \\
\midrule
\multicolumn{2}{@{}l}{\textbf{Synthetic anomalies}}  & & 90 & \\
\midrule
\multicolumn{5}{@{}l}{\textit{Real anomalous classes (test only)}} \\
XRG  & 44 & -- & -- & 22 \\
RRG  & 25 & -- & -- & 13 \\
\midrule
\textbf{Total} & 803 & 584 & 165 & 110 \\
\bottomrule
\end{tabular}
\end{table}

We evaluate our anomaly detection approach using the  FIRST\footnote{Faint Images of the Radio-Sky at Twenty centimeters} Radio Galaxy Anomaly Detection Benchmark (FRGADB) data \cite{becker1995first,brand_2024_13773680}. FRGADB is composed of $\sim\!1000$ samples with four radio galaxy classes:  FRI, FRII, XRG, and RRG, distributed between training, validation and test sets (Table~\ref{original_dataset_distrb}) \cite{fanaroff1974morphology,proctor2011morphological}. The images are of size $150\times150$ pixels with radio sources centered. The images were pre-processed using a morphologically augmented sigma clipping algorithm, which effectively filtered out noise while preserving astronomical source pixels \cite{brand2025cara}.

The presence of XRG and RRG anomalies in both the validation and test sets introduces potential data leakage. This means the model is exposed during validation to patterns that closely resemble those it will encounter during testing, which can lead to an overestimation of its true generalization performance. To address this, we generated 90 synthetic anomalous samples from training images to replace the 22 XRG and 12 RRG galaxies in the given validation set \cite{schluter2022natural}. The new validation set comprises 22 FRI, 53 FRII, and 90 anomalous patterns. The 90 synthetic sources were generated through a systematic process using the training set. First, 10 FRI image pairs were randomly selected without replacement. One image from each pair underwent rotations of 90°, 180°, and 270°, followed by superimposition with its partner using maximum pixel values. This created 30 anomalous sources. The same procedure was applied to 10 FRII pairs and 10 hybrid pairs (FRI-FRII combinations), yielding a total of 90 anomalous images (see examples in the third column of Figure~\ref{fig: example_images}). This procedure ensures that no prior knowledge of the test anomalies is leaked during training or validation, leading to a more reliable assessment of model performance. 

\subsection{Method}

\label{COSFIRE_descriptors}
In this section, we sketch the configuration of COSFIRE filters and the derived feature vectors. A more detailed description can be found in \cite{10.1093/mnras/stae821}.
The responses from a set of COSFIRE filters are employed in a descriptor-based approach to characterize radio sources by their morphological features. This approach is based on the analysis of the shape properties and is automatically configured by applying a series of center-on and center-off Difference-of-Gaussians (DoG) filters to an image. A DoG function approximates the second-order derivative of a Gaussian (Laplacian of Gaussian), offering the advantage of efficient computation due to its separability. These DoG filters come in two variants: center-on filters that detect transitions from bright to dark regions, and center-off filters that identify dark to bright transitions. These filters, varying in scale and polarity (i.e., center-on or center-off), are applied to the input image to produce DoG response maps through convolution operations.

To configure a COSFIRE filter, salient features such as radio source  blobs, jets, or lobes are detected along concentric circles on a given example training image (e.g. a specific radio source type). Keypoints are identified by locating local maxima of the DoG responses along these circles. The number of circles and their respective radii act as hyperparameters of the filter. For each detected keypoint $i$, the following attributes are recorded: distance ($\rho_i$) and angle ($\phi_i$) from the pattern center, filter polarity ($\delta_i$), and the outer Gaussian's standard deviation ($\sigma_i$). These parameters are compiled into a set of tuples, denoted by $C_f$:

\begin{equation}
    C_f = \{(\rho_i, \phi_i, \sigma_i, \delta_i)~|~i = 1,...,n\}
\end{equation}

\noindent where $n$ represents the total detected keypoints. In our data set, radio sources are already centered in the 150$\times$150 images, so we use the image center as the COSFIRE reference point. In principle, one could instead select any point, such as the source’s brightness centroid.

For anomaly detection, COSFIRE filters are first automatically configured to be selective for some randomly selected sources drawn from the training set\footnote{Refer to \cite{10.1093/mnras/stae821} and \cite{10.1093/mnras/staf230} for some examples.}. The configuration procedure defines the selectivity of a COSFIRE filter based on the spatial relationships among the selected keypoints. These configured filters are expected to produce strong responses in images with radio structures similar to those seen during training, and weak responses to unfamiliar patterns, thereby making them suitable to detect anomalies.

The application of a configured COSFIRE filter follows three steps. First, DoG response maps are computed for each unique combination of polarity ($\delta$) and scale ($\sigma$). Second, these response maps are blurred with a Gaussian function to account for positional variations\footnote{See Table~\ref{tab:hyperparameters} for the parameters used to compute the standard deviation $\hat{\sigma}_i$ of the Gaussian, which increases linearly with the distance $\rho_i$ from the filter center. This function introduces spatial tolerance in the filter response. For more details, please refer to \cite{10.1093/mnras/stae821}.}. Without this blurring step, the COSFIRE filter would only respond to exact replicas of the pattern in the training image used for the filter configuration. Therefore, the blurring introduces some tolerance to spatial deviations, enabling the filter to recognize patterns that slightly differ from the preferred configuration. Each blurred response map is then shifted in the opposite direction of its corresponding polar coordinates $(\rho_i, \phi_i)$ so that all DoG responses meet at the filter centre \cite{10.1093/mnras/stae821}. Finally, the COSFIRE filter response is obtained by calculating the geometric mean across all aligned and blurred response maps. Rotational invariance is achieved by generating multiple rotated versions of the base COSFIRE filter, each created by offsetting the angular parameter $\phi_i$ of every keypoint by an angle
\(\psi \in \{0, \tfrac{\pi}{6}, \dots, 11\tfrac{\pi}{6}\}\). A rotation-invariant COSFIRE response is then obtained by taking the maximum response across all these rotated filters. The final COSFIRE descriptor -- a numerical feature vector representing the structural patterns in an image -- is constructed by applying all rotation-invariant filters to an image and recording their maximum responses. This feature vector effectively captures the essential structural elements of a given radio galaxy image. Following \cite{10.1093/mnras/stae821} the descriptors are then L2-normalized. 

\begin{figure*}[t]
    \centering
    \footnotesize
    \includegraphics[trim=0.9cm 1.2cm 0.7cm 0.2cm,clip,width=\textwidth]{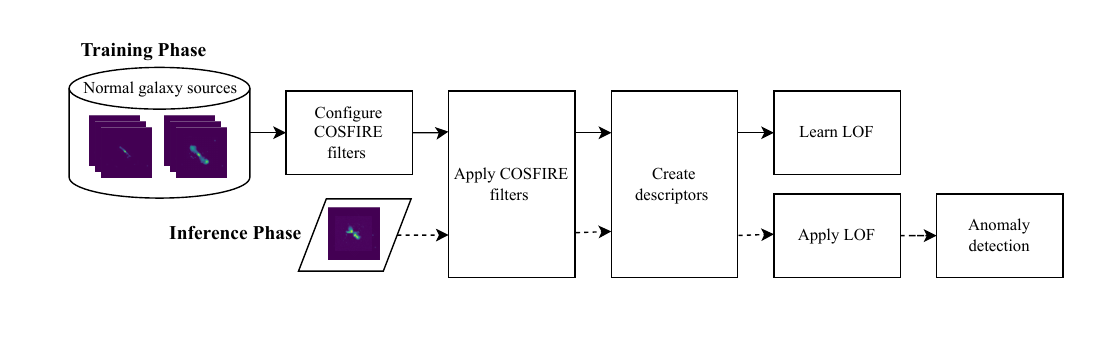}
     \caption{A schematic illustration of the proposed COSFIRE framework for detecting anomalous radio galaxies, showing training and inference phases of the workflow. The training phase comprises three key steps: configuring the COSFIRE filters, extracting descriptors from known radio galaxy images, and training a LOF model for anomaly detection. During the inference phase, new radio images undergo the same descriptor extraction before being evaluated by the trained LOF model to identify potential anomalies. The ``Apply COSFIRE filters" and ``Create descriptors" steps are shared between training and inference stages. The continuous arrows represent the training phase while dashed arrows indicate the inference phase.}
\label{fig:flowchart}
\end{figure*}

We then used the Local Outlier Factor (LOF) algorithm \cite{breunig2000lof} to detect anomalies in COSFIRE descriptors of radio galaxies. The full pipeline is shown in Fig.~\ref{fig:flowchart}.

\begin{table}[t]
\footnotesize
\centering
\caption{Hyperparameter search space in COSFIRE filter optimization. This table outlines the parameter ranges explored to identify optimal filter configurations.}
\label{tab:hyperparameters}
\begin{tabular}{@{}p{0.5\columnwidth}p{0.4\columnwidth}@{}}
\toprule
\textbf{Parameter Name} & \textbf{Values} \\
\midrule 
$\sigma$ – DoG outer standard deviation & $\{5,6\}$ \\
\addlinespace
$P$ – set of concentric circle radii & $\left\{
\begin{array}{l}
\{0,5,10,15,20,25\}, \\
\{0,5,10,15,20,25,30\}
\end{array}
\right\}$ \\
\addlinespace
$t_1$ – response threshold & $\{0.05,0.1\}$\\ 
\addlinespace
\multirow{2}{=}{$\sigma_0^\prime$, $\alpha$ – parameters for positional tolerance} 
& $\sigma_0^\prime \in \{0.5, 0.75\}$ \\
& $\alpha \in \{0.1, 0.15\}$ \\
\addlinespace
\multicolumn{2}{@{}p{\columnwidth}@{}}{
\textit{Note:} The positional tolerance $\hat{\sigma}_i$ for each filter subunit is computed as $\hat{\sigma}_i = \sigma_0^\prime + \alpha \rho_i$, where $\rho_i$ is the radial distance of the subunit from the filter center. This value $\hat{\sigma}_i$ is used as the standard deviation of a Gaussian function that controls the spatial tolerance during filter response computation.}
 \\
\bottomrule
\end{tabular}
\end{table}

\subsection{Evaluation Metrics}
\label{sec:evaluations_metrics}
To evaluate our anomaly detection framework, we employ metrics specifically designed for imbalanced data sets, as anomalous radio galaxy sources represent a small fraction of the data set. We assess model performance using four key metrics: Precision, Recall, Specificity, and G-Mean. Precision quantifies the proportion of true anomalies among all instances predicted as anomalies. Recall measures the model's ability to correctly identify actual anomalous sources. Specificity evaluates the model's accuracy in recognizing normal (non-anamolous) galaxies. G-Mean is the geometric mean of Specificity and Recall, and it reflects the model's balanced performance across both majority and minority classes. These metrics collectively offer a comprehensive assessment of the model's ability to distinguish rare, anomalous radio sources while maintaining reliable detection of typical galaxy morphologies. Note: Performance evaluation considers anomalous sources as positive instances and normal sources as negative instances in all metric calculations.

\section{Experimental Results}
\label{sec: experiments_and_results}

\subsection{Performance}

To evaluate the performance of our COSFIRE filter approach in anomaly detection, we adopted an experimental design closely aligned with the methodologies presented in \cite{10.1093/mnras/stae821} and \cite{10.1093/mnras/staf230}. Training data underwent a stratified hold-out split to ensure representative sampling across classes. This means the dataset was randomly divided into training and validation sets while preserving the original class distribution in both. Prior work \cite{10.1093/mnras/stae821, 10.1093/mnras/staf230} demonstrated that COSFIRE filter-generated descriptors retain robustness across varied hyperparameter spaces. Based on these findings, we simplified our workflow by reducing the hyperparameter search space, selecting two values per COSFIRE hyperparameter instead of three as done in \cite{10.1093/mnras/stae821} (Table~\ref{tab:hyperparameters}).

For our unsupervised learning approach, we optimized the LOF algorithm by systematically exploring its hyperparameters: leaf size $\in$ \{40, 60, 80\}, contamination $\in$ \{0.37, 0.38, 0.39\}, number of neighbors $\in$ \{8, 9, 10\}, and metric $\in$ \{Cosine, Manhattan, Minkowski\} \cite{breunig2000lof}. All remaining parameters were maintained at their default values to constrain the optimization space. Therefore, each COSFIRE configuration underwent 81 distinct experiments during the hyperparameter grid search. 
We performed a full grid search over LOF hyperparameters for each of the 32 COSFIRE configurations, yielding an $81\times32$ matrix of evaluation scores.

We first selected COSFIRE configurations by computing mean G-Mean scores across the validation set. A right-tailed Student's $t$-test was used to compare the highest-performing descriptor against all other configurations \cite{marshall2011introduction}. Descriptors achieving $p$-values $\geq 0.05$ were considered statistically similar to the optimal configuration and retained for further analysis. This filtering identified two descriptors\footnote{They share the same parameter values $\sigma = 6$, $P=\{0,5,10,15,20,25,30\}$, $\sigma_0^\prime=0.75$, and $\alpha=0.15$, and differ only in the value of $t_1$ (\(0.10\) and \(0.05\)).
} that performed consistently well, reducing the evaluation matrix to $81 \times 2$. For each of these descriptors, the top 10 performing LOF hyperparameter sets were selected and evaluated on the test set.

This analysis considered varying numbers of COSFIRE filters, ranging from 5 to 100 in the validation set. As the configuration was performed on two distinct known classes, FRI and FRII (refer to Section ~\ref{COSFIRE_descriptors}), the dimensionality of the feature descriptor scales proportionally with the number of filters, where $n$ filters produced a $2n$-dimensional descriptor (e.g. 5 filters per class generated a 10-dimensional descriptor, 6 filters a 12-dimensional descriptor, and so forth). The experiments revealed that 90 COSFIRE filters per class were required to achieve optimal G-Mean performance on the validation set.

\subsection{Comparison with previous work}

The results of this study are compared with the state-of-the-art findings presented in \cite{brand2025cara}, which used the same data set. Their work employed Principal Component Analysis (PCA) for feature extraction combined with the LOF for anomaly detection. Additionally, they investigated three Convolutional Autoencoder (CAE) architectures: Standard Convolutional Autoencoder (SCAE), Baseline Convolutional Autoencoder (BCAE) \cite{ventura_2022}, and Memory Unit Standard Convolutional Autoencoder (MemSCAE) (see \cite{brand2025cara,gong2019memorizing}). BCAE serves as a shallow network autoencoder that reconstructs input data, using reconstruction error as the loss function. SCAE enhances upon BCAE by applying a deeper network \cite{brand2025cara}. MemSCAE further extends the architecture with a memory unit that constrains the latent space to some degree, thereby improving anomaly detection capabilities \cite{gong2019memorizing}. 

The extensive experiments conducted in \cite{brand2025cara} demonstrated that MemSCAE and BCAE achieved a consistent G-Mean performance of $\sim 0.77$ in anomaly detection tasks, outperforming both PCA + LOF and SCAE approaches. Our COSFIRE + LOF framework surpassed these benchmarks, achieving an average G-Mean of $\sim 0.79$ on the test data set (Table ~\ref{tab:performance_comparison}). Additionally, when using the synthetic validation set (see Section ~\ref{datatab_section}) for hyperparameter tuning, we obtained a G-Mean score of 0.78 $\pm$ 0.03 using 67 COSFIRE filters per class.
 
\begin{table}[t]
\footnotesize
\centering
\caption{Performance comparison of various models using LOF for anomaly detection in a semi-supervised setting. The COSFIRE$^{*}$ model used the synthetic validation set for hyperparameter tuning.}
\begin{tabular}{@{}p{0.2\columnwidth}p{0.21\columnwidth}p{0.21\columnwidth}p{0.21\columnwidth}@{}}
\toprule
\textbf{Model} & \textbf{G-Mean} & \textbf{Specificity} & \textbf{Recall} \\
\midrule
PCA & 0.60 $\pm$ 0.05 & \textbf{0.78 $\pm$ 0.03} & 0.47 $\pm$ 0.07 \\
MemSCAE & 0.77 $\pm$ 0.02 & 0.68 $\pm$ 0.07 & 0.90 $\pm$ 0.06 \\
SCAE & 0.77 $\pm$ 0.01 & 0.69 $\pm$ 0.05 & 0.85 $\pm$ 0.05 \\
BCAE & 0.72 $\pm$ 0.02 & 0.72 $\pm$ 0.09 & 0.73 $\pm$ 0.08 \\
COSFIRE & \textbf{0.79 $\pm$ 0.01} & 0.70 $\pm$ 0.00 & \textbf{0.89 $\pm$ 0.01} \\
COSFIRE$^{*}$ & 0.78 $\pm$ 0.03 & 0.70 $\pm$ 0.02 & 0.88 $\pm$ 0.04\\
\bottomrule
\end{tabular}
\label{tab:performance_comparison}
\end{table}

 \section{Discussion and Conclusion}
 \label{sec:discussion_and_conclusion}

The COSFIRE-based methodology generates robust feature representations that effectively detect anomalies not present in the training data, while maintaining high performance despite the challenges of limited training data and class imbalance. This results in improved performance, with a G-Mean score of 79\%, surpassing both the competing deep learning autoencoder (77\%) and the PCA-based approach (60\%) reported in \cite{brand2025cara}, the latter being a similarly lightweight anomaly detection method.

By leveraging the COSFIRE methodology, our approach addresses the scalability challenge of anomaly detection in next-generation radio surveys by using lightweight, rotation-invariant descriptors that outperform deep autoencoder pipelines on the FIRST benchmark data set. We refer readers to our prior work \cite{10.1093/mnras/stae821,10.1093/mnras/staf230} for detailed FLOP comparisons demonstrating COSFIRE’s efficiency versus deep learning methods. Its demonstrated effectiveness highlights particular promise for enabling serendipitous discoveries with advanced telescopes like the SKA, where efficient processing and the ability to uncover unknown phenomena are crucial. However, the evaluation is limited by the modest size of the FRGADB data set ($\sim1,000$ images) and its focus on only two anomalous classes (XRG/RRG). Future work will expand validation to larger and more diverse catalogs and additional rare morphologies.

\section*{Data and Code Availability}
\noindent The code and data set used in this work are publicly available on GitHub at \url{https://github.com/stevenndungu/anomaly_detection}.
Additionally, a brief preview of the analyses can be found at \url{https://stevenndungu.github.io/anomaly_detection/}.

\section*{Acknowledgements}

\begin{table}[h]
\centering
\footnotesize
\arrayrulecolor{white}
\begin{tabular}{|p{0.95\columnwidth}|}
\hline
Part of this work is supported by the Foundation for Dutch Scientific Research Institutes. \\
\hline
This work is based on the research supported in part by the National Research Foundation of South Africa (grant numbers 119488 and CSRP2204224243). \\
\hline
The financial assistance of the South African Radio Astronomy Observatory (SARAO) towards this research is hereby acknowledged (\url{www.ska.ac.za}). \\
\hline
We thank the Center for Information Technology of the University of Groningen for their support and for providing access to the Hábrók high performance computing cluster. \\
\hline
\end{tabular}
\arrayrulecolor{black}
\end{table}





\begin{table}[h]
\centering
\footnotesize

\arrayrulecolor{white}
\begin{tabular}{|p{0.175\columnwidth}|p{0.75\columnwidth}|}
\hline
\multirow{2}{*}{} & University of Stellenbosch, Computer Science, Cnr Banhoek Road \& Joubert Street, Stellenbosch 7600, South Africa; email: \url{26846578@sun.ac.za}. \\
\cline{2-2}
 \textbf{Steven Ndung'u;}     & University of Groningen, Bernoulli Institute for Mathematics, Computer Science, and Artificial Intelligence, Nijenborgh 9, 9747AG Groningen, The Netherlands; email: \url{s.n.machetho@rug.nl}. \\
\hline
\textbf{Trienko Grobler}; & University of Stellenbosch, Computer Science, Cnr Banhoek Road \& Joubert Street, Stellenbosch 7600, South Africa; email: \url{tlgrobler@sun.ac.za}. \\
\hline
\textbf{Stefan J. Wijnholds}; & ASTRON, Oude Hoogeveensedijk 4, 7991 PD. Dwingeloo, The Netherlands; email: \url{wijnholds@astron.nl}. \\
\hline
\textbf{George Azzopardi}; & University of Groningen, Bernoulli Institute for Mathematics, Computer Science, and Artificial Intelligence, Nijenborgh 9, 9747AG Groningen, The Netherlands; email: url{g.azzopardi@rug.nl}.\\
\hline
\end{tabular}

\arrayrulecolor{black}
\end{table}

\end{document}